\documentclass[journal]{IEEEtran}
\ifCLASSINFOpdf
\else
\fi
\usepackage{amsmath}
\usepackage{amssymb}
\usepackage{graphicx}

\begin{document}
\title{Diffusion-Based Heart Sound Generation: Evaluation with Physiological Signal Metrics, Classifiers, and Expert Listening}

\author{Xinqi Bao, 
        Jia Bi, 
        Xin Chen,
        Ernest Nlandu Kamavuako,
        and~Saikat~Chatterjee 
\thanks{X. Bao is with the Department of Information Science \& Engineering, KTH Royal Institute of Technology, Stockholm, 114 28 Sweden (e-mail: xba@kth.se), and also affiliated with Karolinska Institutet (KI), Stockholm, Sweden (e-mail: xinqi.bao@ki.se).}
\thanks{J. Bi is with Rutherford Appleton Laboratory, United Kingdom.}
\thanks{X. Chen is with Peng Cheng Laboratory, China.}
\thanks{E. N. Kamavuako is with the Department of Engineering, King's College London, United Kingdom.}
\thanks{S. Chatterjee is with the Department of Information Science \& Engineering, KTH Royal Institute of Technology, Stockholm, 114 28 Sweden.}
}

\markboth{}
{Shell \MakeLowercase{\textit{et al.}}: Bare Demo of IEEEtran.cls for IEEE Journals}

\maketitle

\begin{abstract}
Publicly available phonocardiogram (PCG) datasets remain limited in size and pathological diversity, constraining both auscultation training and the generalisation of automated heart-sound classifiers. A class-conditional diffusion model for PCG generation is developed in the log-mel domain and synthetic fidelity is assessed using complementary (i) physiology-inspired plausibility metrics, (ii) downstream label-consistency evaluation, and (iii) expert listening. Experiments use the PhysioNet/Computing in Cardiology Challenge 2016 dataset (3240 recordings) with recording-level splits. After preprocessing and quality control, 16{,}749 non-overlapping 4\,s clips are mapped to a normalised $1\times128\times128$ log-mel representation to train a conditional 2D U-Net denoiser with classifier-free guidance.
Signal-level plausibility is quantified on reconstructed waveforms using three lightweight metrics: an envelope-autocorrelation rhythm score, an amplitude-based explosion score, and the dominant cycle lag. Synthetic clips preserve similar dominant cycle durations but exhibit reduced envelope periodicity and increased transient burstiness relative to real clips. For downstream evaluation, a ResNet-50 classifier achieves 92.24\% accuracy on the held-out real test set and 82.8\% accuracy on class-balanced synthetic batches, indicating that generated signals retain discriminative structure relevant to normal/abnormal classification. In a pilot expert listening study (60 clips, two clinicians), most synthetic clips are judged as heart-sound-like, while abnormality sensitivity is low for both real and synthetic 4\,s excerpts. Overall, the results provide a practical baseline for diffusion-based PCG generation while highlighting remaining challenges in retaining abnormal acoustic cues and reducing reconstruction-induced artefacts.
\end{abstract}

\begin{IEEEkeywords}
phonocardiogram, heart sound, diffusion model, class-conditional generation, log-mel spectrogram, data augmentation, auscultation
\end{IEEEkeywords}

\IEEEpeerreviewmaketitle

\section{Introduction}

Cardiovascular diseases (CVDs) remain the leading cause of death worldwide, with the World Health Organization estimating 19.8 million deaths in 2022 (approximately 32\% of all global deaths) \cite{who_cvd_2025}. In this context, cardiac auscultation remains a widely used, low-cost triage tool in primary care and community settings, yet diagnostic performance for valvular heart disease varies substantially across studies, lesions, and care levels \cite{davidsen2023ausc}. A more recent study further highlighted pronounced inter-clinician variability and consistently better performance among cardiologists compared with non-specialists \cite{gunasekera2025ausc}. Collectively, these findings suggest that auscultation accuracy remains strongly dependent on experience and local training.

Data scarcity is a central bottleneck for both clinical education and automated analysis. Medical students and junior doctors have limited access to large, diverse phonocardiogram (PCG) libraries that cover multiple devices, recording environments, and lesion types during time-constrained rotations. In parallel, automatic heart sound classification has advanced substantially \cite{prince2023murmur}. The PhysioNet/Computing in Cardiology Challenge 2016 released over 3000 multi-institutional PCG recordings for normal-versus-abnormal classification and established a widely used benchmark \cite{liu2016cinc}. More recently, the CirCor DigiScope dataset extended this line of work to murmur detection and multi-class murmur classification in paediatric populations \cite{oliveira2022circor}. Despite these efforts, the number and diversity of publicly available, well-annotated PCG datasets remain limited, which constrains the generalisation of machine learning models across hospitals, devices, and case-mix profiles.

Generative modelling offers a principled approach to mitigate data scarcity and improve robustness for physiological signals. For heart sounds, Narv\'aez and Percybrooks synthesised PCG segments using generative adversarial networks with empirical wavelet transform-based post-processing and reported acoustic similarity to real recordings \cite{narvaez2020ganpcg}. However, existing work has largely focused on normal sounds and provides limited evidence regarding pathological diversity, rhythmic structure, or downstream utility \cite{abbott2026syntheticaudio}. Diffusion models have recently emerged as a strong alternative, with favourable training stability and flexible conditional generation. BioDiffusion demonstrated high-fidelity synthesis of multivariate biomedical signals and illustrated benefits for class imbalance and limited labels across multiple modalities \cite{li2024biodiffusion}. In electrocardiography, diffusion-based approaches have been explored for ECG synthesis and for improving label efficiency and personalisation in arrhythmia detection \cite{neifar2023diffecg,zhou2025diffecg}.

In contrast, diffusion-based generation for heart sounds remains relatively underexplored, and there is no widely adopted quantitative framework to assess whether synthetic PCG segments preserve clinically meaningful S1--S2 rhythm structure and energy distribution while avoiding physiologically implausible artefacts under realistic noise and device conditions. Existing evaluations often rely on visual inspection or downstream task performance alone, which only partially reflects clinical plausibility.

This paper addresses part of this gap with three contributions. First, a conditional diffusion framework tailored to heart sounds is introduced by training a compact 2D U-Net on log-mel representations of PCG segments, using normal/abnormal labels for controllable generation. Second, a set of physiology-inspired plausibility metrics is introduced to constrain the generated signals from complementary physical perspectives, including cardiac cycle duration, rhythm stability, and waveform physical feasibility, operationalised via an envelope-autocorrelation rhythm score, an amplitude-based explosion score, and the corresponding dominant cycle lag. These metrics are lightweight proxy measures of signal plausibility and are not direct
physiological measurements or clinical biomarkers. Third, these metrics are linked to downstream and human-level evaluation via (i) a supervised log-mel classifier assessing normal/abnormal discriminability and (ii) a small-scale expert listening study in which clinicians identify murmurs and judge whether segments are real or synthetic. The overarching goal is to establish a practical baseline for diffusion-based PCG generation that balances perceptual plausibility with statistical and physiological credibility on open datasets.

\section{Methods}

\subsection{Dataset and Preprocessing}
\label{subsec:dataset}
Recordings from the PhysioNet/Computing in Cardiology Challenge 2016 dataset were used \cite{liu2016cinc}. The dataset contains 3240 recordings with expert record-level labels (normal vs.\ abnormal): 2575 normal (79.5\%) and 665 abnormal (20.5\%).

To prevent leakage between data partitions, the data split was performed at the recording level prior to any segmentation. Recordings were stratified by label and split into training, validation, and test sets with approximate proportions of 70\%, 15\%, and 15\%. All segments derived from the same original recording were restricted to the same subset.

Each recording was converted to a single-channel waveform at the native sampling rate (2000~Hz). A constant offset was removed by subtracting the recording-wise mean, followed by a 20--500~Hz band-pass filter to suppress very low-frequency drift and high-frequency noise while retaining the dominant heart-sound energy. Filtered waveforms were segmented into non-overlapping 4~s clips, and each clip was peak-normalised by its maximum absolute amplitude \cite{bao2022duration}.

Quality control excluded clips with very low energy or substantial clipping. Specifically, clips with root-mean-square (RMS) amplitude below $10^{-3}$, or with more than 1\% of samples exceeding $|0.99|$ after normalisation, were discarded. This preprocessing yielded 16{,}749 usable 4~s clips in total (12{,}827 normal; 3{,}922 abnormal; 23.4\% abnormal). These clips and their labels were used as the sole inputs for both diffusion training and downstream classification. Segment counts by split are reported in Table~\ref{tab:dataset_split}.

\subsection{Log-mel Representation}
\label{subsec:logmel}
Log-mel features are widely used in PCG classification and provide a compact 2D domain that can be shared by the diffusion generator and the downstream classifier. Each quality-controlled 4 s PCG clip was converted to a log-mel time--frequency representation. 

To obtain a fixed 128×128 time–frequency grid as network input, clips were right-padded to 8128 samples prior to feature extraction. A short-time Fourier transform (STFT) was then computed using a 512-point FFT with a 512-sample Hann window, hop size 64, and centered framing with reflection padding. The resulting spectra were converted to a power representation and projected onto a 128-bin mel filterbank spanning 20--500 Hz. Mel values were floored to a small positive constant $\epsilon_{\mathrm{mel}}$ to avoid numerical issues and then converted to the natural logarithm domain, producing a $128 \times 128$ log-mel tensor (frequency $\times$ time).

To ensure a consistent feature scale across the diffusion model and classifier, global standardization statistics were estimated from a random subset of the training set and then applied unchanged to validation and test data. Specifically, 256 training clips were sampled with a fixed random seed (seed=0), and the global mean and standard deviation were computed over all mel bins and time frames by aggregating the sum and squared-sum of log-mel values. The final normalized representation had shape $1 \times 128 \times 128$ (channel $\times$ frequency $\times$ time) and was used consistently for diffusion training/sampling and classification experiments.

\begin{table}[t]
\caption{PCG segment counts and class composition after preprocessing (split performed at the recording level).}
\label{tab:dataset_split}
\centering
\begin{tabular}{lrrr}
\hline
Split & Segments & Normal, n (\%) & Abnormal, n (\%) \\
\hline
Train      & 11785 & 8994 (76.3)  & 2791 (23.7) \\
Validation & 2489  & 1898 (76.3)  & 591  (23.7) \\
Test       & 2475  & 1935 (78.2)  & 540  (21.8) \\
\hline
Total      & 16749 & 12827 (76.6) & 3922 (23.4) \\
\hline
\end{tabular}
\end{table}

\begin{figure*}[t]
  \centering
  \includegraphics[width=\textwidth]{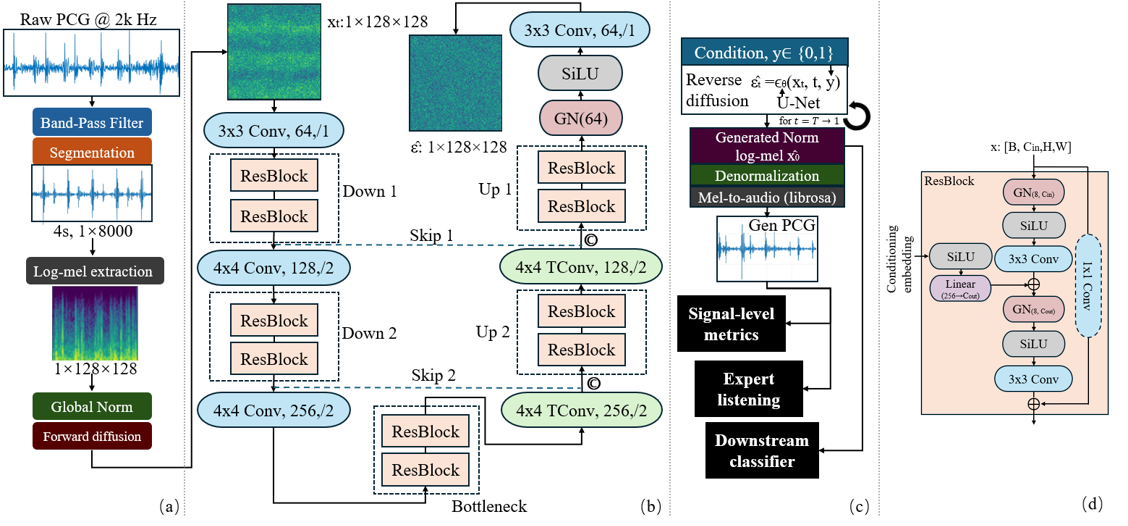}
  \caption{Overview of the proposed pipeline. (a) Preprocessing and log-mel feature extraction. (b) Compact 2D U-Net denoiser for reverse diffusion with skip connections. (c) Conditional generation and downstream evaluation; the reverse diffusion block is shown schematically. (d) Residual block with time/class conditioning injection.}
  \label{fig:pipeline}
\end{figure*}

\subsection{Diffusion-based PCG Generator}
\label{subsec:diffusion}

Figure~\ref{fig:pipeline} summarises the proposed diffusion-based PCG generator and evaluation pipeline. The generator operates on the normalised log-mel representation
$\tilde{\mathbf{X}} \in \mathbb{R}^{1 \times 128 \times 128}$ and is trained as a conditional denoising diffusion model with a compact 2D U-Net backbone. Conditioning is provided by the binary class label $y \in \{0,1\}$, indicating normal versus abnormal segments. Classifier-free guidance (CFG) is applied at sampling time to enable both conditional and approximately unconditional generation.

\subsubsection{Forward diffusion process}
\label{subsubsec:forward_diffusion}

A standard discrete-time diffusion process with $T=1000$ steps is defined in the log-mel domain. Let $\mathbf{x}_0 \equiv \tilde{\mathbf{X}}$ denote a clean normalised log-mel tensor. The noise schedule is specified by
$\beta_t \in (0,1)$, with $\beta_t$ linearly increasing from $10^{-4}$ to $2\times 10^{-2}$ across $t=1,\ldots,T$.
Define $\alpha_t = 1-\beta_t$ and $\bar{\alpha}_t = \prod_{s=1}^{t}\alpha_s$.
The forward process adds Gaussian noise as
\[
q(\mathbf{x}_t \mid \mathbf{x}_0) = \mathcal{N}\!\left(\mathbf{x}_t;\, \sqrt{\bar{\alpha}_t}\,\mathbf{x}_0,\,(1-\bar{\alpha}_t)\mathbf{I}\right),
\]
equivalently constructed by sampling ${\epsilon}\sim\mathcal{N}(\mathbf{0},\mathbf{I})$ and setting
\[
\mathbf{x}_t = \sqrt{\bar{\alpha}_t}\,\mathbf{x}_0 + \sqrt{1-\bar{\alpha}_t}\,{\epsilon}.
\]
During training, a timestep $t$ is sampled uniformly from $\{1,\ldots,T\}$ for each example.

\subsubsection{Denoiser architecture and training}
\label{subsubsec:unet_training}

The denoiser ${\epsilon}_\theta(\mathbf{x}_t, t, \tilde{y})$ predicts the added noise ${\epsilon}$ from a noisy input $\mathbf{x}_t$. The network input has shape $[B,1,128,128]$ (batch, channel, frequency, time). A $3\times3$ convolution maps the single-channel input to 64 feature maps. The timestep $t$ is embedded using sinusoidal features followed by a two-layer MLP (hidden width 256) to obtain a 256-dimensional vector; the class label is mapped to a 256-dimensional vector via a learned embedding. These two vectors are summed to form a single conditioning vector, which is injected into each residual block via a linear projection and channel-wise addition (Figure~\ref{fig:pipeline}b,d).

The U-Net comprises two downsampling stages (64$\rightarrow$128$\rightarrow$256 channels), a bottleneck, and two upsampling stages with skip connections. Each stage uses residual blocks with GroupNorm (8 groups) and SiLU activations. Downsampling is performed using $4\times4$ stride-2 convolutions and upsampling with $4\times4$ stride-2 transposed convolutions. A final GroupNorm and $3\times3$ convolution produce a single-channel noise prediction with the same spatial shape as the input, interpreted as ${\epsilon}_\theta(\mathbf{x}_t,t,\tilde{y})$.

CFG is enabled by label dropout: during training, the true label $y \in \{0,1\}$ is replaced by a special ``null'' label $\varnothing$ with probability $P_{\mathrm{uncond}}=0.10$, yielding $\tilde{y}\in\{0,1,\varnothing\}$. The model is trained with the standard noise-prediction objective
\[
\mathcal{L}(\theta)=\mathbb{E}\left[\left\lVert {\epsilon}-{\epsilon}_\theta(\mathbf{x}_t,t,\tilde{y})\right\rVert_2^2\right],
\]
where the expectation is over $(\mathbf{x}_0,y)$ from the training set, $t$ sampled uniformly, and ${\epsilon}\sim\mathcal{N}(\mathbf{0},\mathbf{I})$.

Optimisation uses AdamW with learning rate $2\times10^{-4}$, weight decay $10^{-4}$, batch size 64, and gradient-norm clipping at 1.0, for 10{,}000 update steps on the training split (11{,}785 segments). An exponential moving average (EMA) of parameters with decay 0.999 is maintained during training and used for all sampling and downstream evaluations.

\subsubsection{Sampling and classifier-free guidance}
\label{subsubsec:sampling_cfg}

Sampling is performed with a deterministic implicit sampler (DDIM; Denoising Diffusion Implicit Models)~\cite{song2020ddim} using 100 steps (i.e., $\eta=0$), subsampling timesteps from the $T=1000$ training schedule. Starting from $\mathbf{x}_T\sim\mathcal{N}(\mathbf{0},\mathbf{I})$, the sampler iteratively produces $\mathbf{x}_{t-1}$ from $\mathbf{x}_t$ using the EMA model predictions. In Figure~\ref{fig:pipeline}c, reverse diffusion is illustrated schematically as a conditional denoiser $\hat{\epsilon}={\epsilon}_\theta(\mathbf{x}_t,t,\tilde{y})$; in practice, CFG combines conditional and unconditional predictions as
\[
\widehat{{\epsilon}}(\mathbf{x}_t,t,y) = (1+w)\,{\epsilon}_\theta(\mathbf{x}_t,t,y) - w\,{\epsilon}_\theta(\mathbf{x}_t,t,\varnothing),
\]
with guidance scale $w=1.2$. The generated log-mel samples are subsequently used for quantitative evaluation, classifier training/testing, and waveform reconstruction.

\subsection{Physiology-inspired Plausibility Metrics}
\label{subsec:metrics}

To obtain a simple yet physically interpretable view of how synthetic PCG segments differ from real recordings, three scalar metrics were computed in the time domain.  
Rather than serving as clinical biomarkers, these metrics are designed to constrain the generated signals from complementary physical perspectives:  
(i) whether the implied cardiac cycle duration lies in a physiologically plausible range,  
(ii) whether this periodic structure is temporally stable across successive cycles, and  
(iii) whether the waveform satisfies the physical constraints of a continuous, smooth, finite-bandwidth mechanical vibration.  
All metrics were evaluated on 4~s waveforms after the same 20--500~Hz band-pass filtering and normalisation described in preprocessing.  
The metrics are inexpensive, reproducible, and agnostic to the specific generative model.

\subsubsection{Rhythm score (envelope autocorrelation peak)}
\label{subsubsec:rhythm}

Let $s[n]$ denote a PCG segment of length $N$ sampled at rate $f_s$.  
An amplitude envelope $e[n]$ was obtained as the magnitude of the analytic signal,
\begin{equation}
e[n] = \bigl|\mathcal{H}\{s[n]\}\bigr|,
\end{equation}
followed by mean subtraction $e[n]\leftarrow e[n]-\frac{1}{N}\sum_{m=0}^{N-1}e[m]$ to remove the DC component.  
This envelope captures the temporal pattern of mechanical energy release associated with successive S1--S2 events.

To quantify the stability of this cardiac cycle pattern, the (biased) autocorrelation of $e[n]$ was computed
\begin{equation}
R_e[k] = \sum_{n=0}^{N-k-1} e[n]\,e[n+k],
\end{equation}
and normalised by the zero-lag value
\begin{equation}
r[k] = \frac{R_e[k]}{R_e[0]+\varepsilon},
\end{equation}
where $k$ is the lag in samples and $\varepsilon$ ($10^{-8}$) is a small constant for numerical stability.  
The search was restricted to physiologically plausible heart periods by considering
\begin{equation}
L_{\min} = \left\lceil 0.33\, f_s \right\rceil, 
\quad
L_{\max} = \left\lfloor 1.50\, f_s \right\rfloor ,
\end{equation}
corresponding approximately to 40--180 beats per minute.  
The rhythm score was defined as
\begin{equation}
\text{rhythm\ score} = \max_{k \in [L_{\min}, L_{\max}]} r[k].
\end{equation}

A high rhythm score indicates that the energy envelope of the signal exhibits a strong and temporally consistent repetition pattern across successive cardiac cycles, whereas irregular, weakly periodic, or noise-dominated segments yield lower values.

\subsubsection{Explosion score (transient amplitude ratio)}
\label{subsubsec:explosion}

While RMS normalisation equalises overall energy, unstable synthesis or reconstruction can introduce extreme, short-lived transients that violate the physical constraints of cardiac mechanical vibration.  
To capture such behaviour, the maximum and median absolute amplitudes were computed:
\begin{equation}
a_{\max} = \max_{0\le n < N} |s[n]|,
\qquad
a_{\mathrm{med}} = \operatorname{median}_{0\le n < N} |s[n]|.
\end{equation}
The explosion score was defined as
\begin{equation}
\text{explosion\ score} = \frac{a_{\max}}{a_{\mathrm{med}} + \varepsilon},
\end{equation}
with $\varepsilon$ for numerical stability.  
Large values indicate the extreme transient events whose amplitude is inconsistent with the typical vibration level of the signal, and are therefore suggestive of non-physiological artefacts such as impulsive noise, clipping, or unstable waveform reconstruction.

\subsubsection{Dominant cycle lag (cardiac period estimate)}
\label{subsubsec:lag}

Finally, the lag at which the envelope autocorrelation attains its maximum within the heart-rate range was recorded as
\begin{equation}
k^{\ast} = \arg\max_{k \in [L_{\min}, L_{\max}]} r[k],
\end{equation}
and converted to seconds:
\begin{equation}
\text{best\ peak\ lag} = \frac{k^{\ast}}{f_s}.
\end{equation}
This quantity provides a coarse estimate of the dominant cardiac cycle duration (and thus heart rate) implied by the signal.  
Values far outside typical adult ranges (approximately 0.6--1.0~s at rest) indicate implausible repetition patterns or artefact-dominated signals.

\subsection{Classifier for Downstream Evaluation}
\label{subsec:classifier}

To assess whether generated PCG segments are useful for downstream tasks, a supervised segment-level classifier was trained on the same normalised log-mel representation as the diffusion generator.

\subsubsection{Architecture and input}
\label{subsubsec:clf_arch}

Each 4~s segment is mapped to a single-channel log-mel tensor as in Section~\ref{subsec:logmel}, with size $1 \times 128 \times 128$. The same global statistics $(\mu_{\mathrm{logmel}}, \sigma_{\mathrm{logmel}})$ estimated from the training split are reused to z-score all segments.

A 2D residual CNN following the overall ResNet-50 design is used as backbone \cite{bao2023time}. The first convolution is adapted to a single input channel, and the final layer is replaced by a two-unit softmax head for \emph{normal} ($y{=}0$) vs.\ \emph{abnormal} ($y{=}1$) classification. Similar time--frequency CNN backbones have been effective for PCG murmur detection \cite{xu2022hierarchical,walker2022dualbayesianresnet}.

\subsubsection{Training protocol}
\label{subsubsec:clf_train}

The classifier is trained only on real segments using the fixed train/validation/test split defined in Section~\ref{subsec:dataset}. To mitigate class imbalance, a weighted cross-entropy loss is used with inverse-frequency class weights, where the weight for each class is set to the reciprocal of its number of training samples. Optimisation uses mini-batch AdamW (lr $10^{-4}$, weight decay $10^{-4}$) for 5 epochs without a learning-rate scheduler, selecting the final checkpoint by validation accuracy. All test-set metrics are computed once on the held-out test split and are not used for model selection or hyperparameter tuning.

\subsubsection{Use for evaluating synthetic PCG}
\label{subsubsec:clf_use}

The trained classifier is used to (i) report a real-test baseline, (ii) evaluate the listening-study clip set, and (iii) stress-test larger uncurated synthetic batches to estimate the fraction of generated samples consistent with the target label. These task-level results complement the signal-level metrics (Section~\ref{subsec:metrics}) and expert listening outcomes (Section~\ref{subsec:expert}).

\subsection{Expert Listening Study}
\label{subsec:expert}

A small expert listening study was conducted to complement signal-level and classifier-based evaluations.

\subsubsection{Stimuli and tasks}
\label{subsubsec:expert_stimuli}

Sixty 4~s audio clips were assembled: 40 real (20 normal, 20 abnormal) and 20 synthetic (10 conditioned on $y{=}0$, 10 on $y{=}1$). Real clips were sampled from the dataset in Section~\ref{subsec:dataset}; synthetic clips were generated by the proposed diffusion model after training. All clips were peak-normalised to a fixed level and exported as de-identified audio files. Raters were informed that both real and generated clips were included, but not the proportions. For each clip, two binary judgements were collected: (i) \emph{normal} vs.\ \emph{abnormal} and (ii) \emph{real} vs.\ \emph{synthetic}.

\subsubsection{Procedure and analysis}
\label{subsubsec:expert_proc}

Two cardiologists completed the task independently and could replay clips before finalising decisions. For the normal/abnormal task, accuracy and confusion matrices are reported separately for real and synthetic clips. For the real/synthetic task, detection and false-alarm rates are reported for synthetic and real clips, respectively, alongside inter-rater agreement.

\section{Results}

\subsection{Qualitative Comparison: Real vs.\ Synthetic Examples}
\label{subsec:qual}

\begin{figure*}[t]
  \centering
  \includegraphics[width=0.9\textwidth]{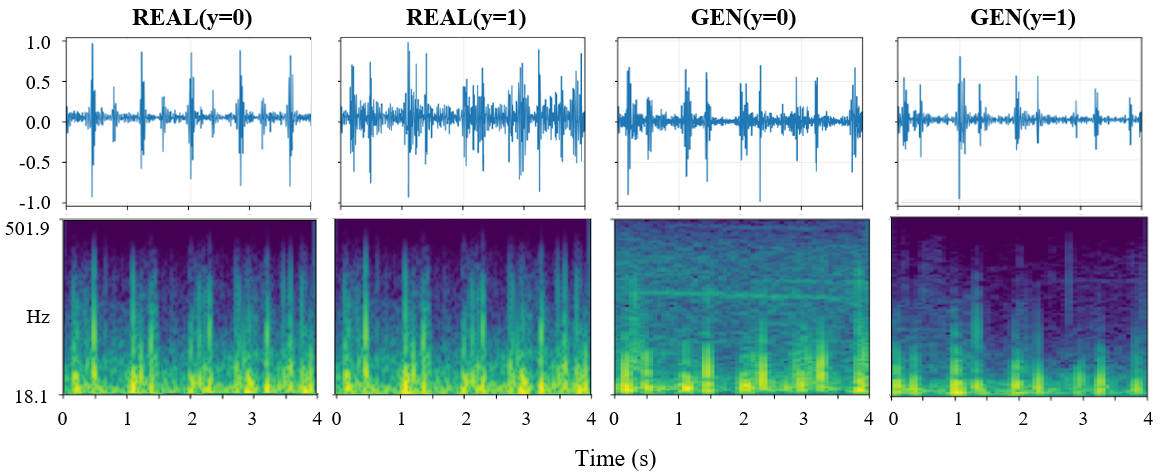}
  \caption{Qualitative examples of real and generated PCG segments (4~s). Top row: waveforms. Bottom row: log-mel representations using 128 mel bands spanning 20--500~Hz (mel-frequency bins shown on the vertical axis).}
  \label{fig:qual_examples}
\end{figure*}

Figure~\ref{fig:qual_examples} shows representative 4~s PCG segments from real recordings and diffusion-generated samples for both classes (normal: $y{=}0$, abnormal: $y{=}1$). For each example, the time-domain waveform (top) and the corresponding log-mel representation (bottom; computed as in Section~\ref{subsec:logmel}) are displayed.
Overall, the generated samples exhibit transient events and low-frequency energy patterns that qualitatively resemble heart-sound structure observed in real recordings. However, some synthetic examples appear spectro-temporally smoother and less variable than their real counterparts, motivating the quantitative evaluations in the following subsections.

\subsection{Signal-level Comparison Using Plausibility Metrics}
\label{subsec:results_metrics}

Table~\ref{tab:metrics_real_vs_gen} reports the three plausibility metrics defined in Section~\ref{subsec:metrics} for real and diffusion-generated PCG clips (4~s; 10 per class), summarised as median [Q1, Q3]. Compared with real clips, generated clips show a lower rhythm score (0.368 vs.\ 0.460), indicating weaker envelope periodicity within the physiological heart-cycle range. In contrast, the explosion score is higher for generated clips (39.00 vs.\ 31.24), consistent with more pronounced transient burstiness. The best peak lag remains similar between real and generated clips (0.845~s vs.\ 0.802~s), suggesting that the dominant cycle duration is broadly preserved despite reduced rhythmic regularity.

\begin{table}[t]
\centering
\caption{Plausibility metric comparison between real and diffusion-generated PCG clips (4~s each; 10 per class). Values are reported as median [Q1, Q3].}
\label{tab:metrics_real_vs_gen}
\renewcommand{\arraystretch}{1.15}
\begin{tabular}{lcc}
\hline
\textbf{Metric} & \textbf{Real} & \textbf{Synthetic} \\
\hline
Rhythm score            & 0.460 [0.449, 0.495] & 0.368 [0.318, 0.413] \\
Explosion score         & 31.24 [26.61, 37.46] & 39.00 [33.04, 45.04] \\
Best peak lag (s)       & 0.802 [0.774, 0.879] & 0.845 [0.768, 0.897] \\
\hline
\end{tabular}
\end{table}

\subsection{Downstream Classification Evaluation}
\label{subsec:results_classifier}

\begin{figure}[t]
\centering
\includegraphics[width=0.85\linewidth]{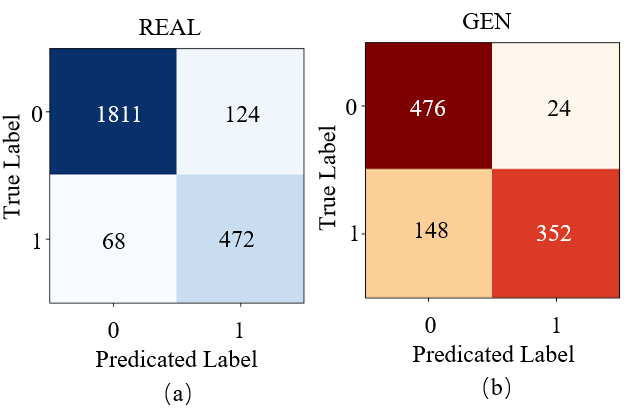}
\caption{Confusion matrices of the ResNet-50 classifier on (a) real test segments and (b) 1000 unfiltered synthetic segments (500 per class).}
\label{fig:cm_real_gen}
\end{figure}

Figure~\ref{fig:cm_real_gen} summarises the ResNet-50 classifier performance on (a) the held-out real test set and (b) unfiltered synthetic segments generated by the diffusion model (1000 segments, 500 per class).
On the real test set ($n=2475$), the classifier achieved an accuracy of 92.24\% (2283/2475), with class-wise recall of 93.6\% for $y{=}0$ (1811/1935) and 87.4\% for $y{=}1$ (472/540), indicating strong segment-level normal-versus-abnormal discrimination under the fixed split.

When evaluated on the unfiltered synthetic set, the overall accuracy decreased to 82.8\% (828/1000). Class-$0$ synthetic samples remained largely separable (476/500 correct; recall 95.2\%), whereas class-$1$ was more frequently mapped to class $0$ (352/500 correct; recall 70.4\%). This asymmetric degradation suggests that, without additional constraints, a subset of generated abnormal segments exhibits log-mel characteristics that overlap with the classifier's normal region.

\subsection{Expert Listening Study}
\label{subsec:results_expert}

Two clinicians independently evaluated 60 de-identified 4~s clips (40 real, 20 synthetic; both balanced by label). Raters were informed that synthetic clips were included, but not the proportion. Each clip received two binary judgements: (i) the dataset label ($y\in\{0,1\}$; normal vs.\ abnormal) and (ii) perceptual plausibility (heart-sound-like vs.\ obvious non-physiological artefact).

Plausibility ratings were high for both sources. For real clips, both clinicians marked 38/40 as plausible (95\%). For synthetic clips, Clinician~A marked 17/20 (85\%) and Clinician~B marked 16/20 (80\%) as plausible (Table~\ref{tab:expert_listening_summary}).

For label discrimination, both clinicians exhibited high specificity but low recall on both real and synthetic subsets. On real clips, Clinician~A achieved 24/40 accuracy (recall 5/20; specificity 19/20) and Clinician~B achieved 21/40 (recall 3/20; specificity 18/20). On synthetic clips, Clinician~A achieved 12/20 (recall 2/10; specificity 10/10) and Clinician~B achieved 11/20 (recall 1/10; specificity 10/10).

\begin{table}[t]
\centering
\caption{Expert listening summary. ``Plausible PCG'' denotes clips judged as heart-sound-like without obvious non-physiological artefacts. Label-discrimination metrics are computed against the dataset label ($y\in\{0,1\}$).}
\label{tab:expert_listening_summary}
\renewcommand{\arraystretch}{1.15}
\begin{tabular}{llcc}
\hline
\textbf{Metric} & \textbf{Clinician} & \textbf{Real (N=40)} & \textbf{Gen (N=20)} \\
\hline
Plausible PCG (rate)
& A & 38/40 (95\%) & 17/20 (85\%) \\
& B & 38/40 (95\%) & 16/20 (80\%) \\
\hline
Accuracy
& A & 24/40 (60\%) & 12/20 (60\%) \\
& B & 21/40 (52.5\%) & 11/20 (55\%) \\
\hline
Recall ($y=1$)
& A & 5/20 (25\%) & 2/10 (20\%) \\
& B & 3/20 (15\%) & 1/10 (10\%) \\
\hline
Specificity ($y=0$)
& A & 19/20 (95\%) & 10/10 (100\%) \\
& B & 18/20 (90\%) & 10/10 (100\%) \\
\hline
\end{tabular}
\end{table}

\section{Discussion}

\subsection{Summary of findings}

Across qualitative, signal-level, downstream, and clinician evaluations, diffusion-generated PCG segments preserved a broadly plausible dominant cycle duration (best peak lag), but showed reduced envelope periodicity (lower rhythm score) and increased transient burstiness (higher explosion score) relative to real clips. These findings support the interpretation of the proposed metrics as complementary plausibility constraints on cardiac cycle duration, rhythm stability, and waveform physical feasibility, rather than as purely statistical features.

The downstream ResNet classifier remained strong on real test data, whereas performance degraded on unfiltered synthetic batches, particularly for the $y{=}1$ class, suggesting class-dependent overlap between generated $y{=}1$ samples and the classifier's normal decision region. Importantly, in this work the classifier is used as a probe of label-consistency rather than as evidence that synthetic data improves real-world generalisation. 
In the pilot listening study, clinicians rated most clips as heart-sound-like, but showed high specificity and low recall for $y{=}1$ on both real and synthetic subsets.

\subsection{Interpretation and sources of error}
Several factors may explain these patterns. First, the ground-truth labels in the PhysioNet/CinC 2016 dataset are defined at the recording level (normal vs.\ abnormal, with limited diagnostic granularity), whereas our evaluation is performed on 4~s segments; pathological evidence such as murmurs or other abnormal cues can be intermittent and may not be present in every short excerpt, reducing both human and model sensitivity \cite{liu2016cinc}. More broadly, acquisition hardware and recording conditions can induce systematic shifts in PCG signal statistics, which may further reduce cross-setting comparability for both perceptual and model-based assessments \cite{bao2024signal}. Second, clinicians reported limited confidence when judging abnormality from 4~s clips: short context can obscure rhythm consistency and non-sustained murmurs, even when replay is allowed. Relatedly, peak normalisation removes amplitude cues that may otherwise support human judgements under variable acquisition conditions, potentially promoting conservative (normal-leaning) decisions in ambiguous cases. Third, raters were informed that synthetic clips were included; in a small-sample setting this may introduce expectation effects in borderline cases, influencing perceptual plausibility ratings (\emph{heart-sound-like} vs.\ \emph{artefact}).

\subsection{Metric-based curation and its trade-off}
Metric-based selection using rhythm, explosion, and dominant-lag criteria can produce curated clips that sound plausible and exhibit physiologically reasonable timing. However, these metrics primarily constrain generic signal plausibility rather than diagnostic semantics. In particular, for the abnormal class, stronger plausibility constraints may inadvertently suppress or truncate discriminative pathological cues (e.g., low-SNR, non-stationary murmur components), thereby biasing selected synthetic clips toward the normal decision region. This highlights that post-hoc physiological filtering is best viewed as a quality-control tool (e.g., for assembling listening stimuli) rather than a substitute for class-consistent generative modelling.

\subsection{Future directions}
Several extensions may improve both plausibility and label consistency. A natural next step is multi-task training, for example by adding an auxiliary classification head or incorporating a pretrained classifier-based consistency loss, thereby jointly optimising denoising fidelity and class discriminability \cite{dhariwal2021diffusionbeatgans,ho2022cfg}. In addition, physiological metrics could inform training through carefully designed, differentiable proxy losses (e.g., envelope periodicity surrogates or transient-penalty regularisers), as direct optimisation of non-differentiable heuristics is non-trivial. Finally, generating longer-context segments may better capture multi-cycle rhythm and sustained abnormal patterns important for clinician assessment, although this introduces a compute--context trade-off; hierarchical or patch-based generation could provide a practical compromise. Beyond binary labels, extending to murmur datasets with richer annotations, such as CirCor DigiScope and related PhysioNet Challenge 2022 resources, is promising but will require careful handling of class imbalance and domain shift \cite{oliveira2022circor,reyna2023plos_physionet2022}.

\subsection{Limitations}
The listening study involved only two clinicians and a limited number of clips, and should be interpreted as a pilot sanity check rather than a powered clinical validation. Short 4~s segments and recording-level labels likely limit achievable sensitivity for both human and automated assessments. Moreover, computed metrics after waveform reconstruction may be influenced by inversion artefacts, and transient artefacts remain a failure mode in a subset of generated samples.

\section{Conclusion}
This work presented a conditional diffusion-based framework for generating PCG segments in the log-mel domain and assessed perceptual and physiological plausibility using complementary signal-level, downstream, and clinician-centred evaluations. Across analyses, generated samples preserved a broadly plausible dominant cycle duration, but exhibited weaker envelope periodicity and more transient artefacts than real clips. Consistent with these deviations, downstream classification performance degraded on unfiltered synthetic batches, with a pronounced drop in abnormal ($y{=}1$) recall. A pilot listening study found high perceived plausibility but low sensitivity for abnormality on short 4~s clips.

Overall, the results indicate that diffusion models can produce heart-sound-like signals with realistic timing, while highlighting current limitations in pathological semantic consistency and artefact control. Future work should prioritise objectives that jointly constrain plausibility and label-consistency (e.g., multi-task or classifier-consistency losses), explore longer-context generation to better capture sustained abnormal patterns, and validate findings with larger, blinded clinician studies and richer multi-class datasets.

\section*{Ethics Statement}
This study used publicly available, de-identified phonocardiogram recordings from the PhysioNet/Computing in Cardiology Challenge 2016 dataset. No new human-subject data were collected in this work.





\ifCLASSOPTIONcaptionsoff
  \newpage
\fi

\end{document}